\documentclass{aastex}
\usepackage{emulateapj5}
\usepackage{apjfonts}
\usepackage{epsf}

\slugcomment{Submitted to the Astrophysical Journal Letters}
\shorttitle{MASS DISTRIBUTION OF BLACK HOLES IN BL LACS}
\shortauthors{Wang et al. \ 2001}

\def\bhm{M_{\rm BH}}

\def\cd{{\cal D}}
\def\cf{{\cal F}}
\def\chii{\chi_{_1}}
\def\chiii{\chi_{_2}}
\def\dotm{\dot{M}}
\def\ga{\gamma}
\def\hb{H$\beta$~}
\def\kes{\kappa_{\rm es}}
\def\lbz{L_{\rm BZ}}
\def\lpk{L_{\rm pk}}
\def\npkp{\nu^{\prime}_{\rm pk}}
\def\nupk{\nu_{\rm pk}}
\def\pehp{\langle\epsilon^{\rm H}_{\rm r}\rangle}
\def\pelp{\langle\epsilon^{\rm L}_{\rm r}\rangle}

\def\sunm{M_{\odot}}

\def\xpk{x_{\rm pk}}

\begin{document}

\title{The Mass Distribution of Black Holes in BL Lacertae Objects} 

\author{Jian-Min Wang\altaffilmark{1}, 
Sui-Jian Xue\altaffilmark{2},
and Jian-Cheng Wang\altaffilmark{3}}

\altaffiltext{1}{
Laboratory for High Energy Astrophysics,
Institute of High Energy Physics, CAS,
Beijing 100039, P.R. China, wangjm@astrosv1.ihep.ac.cn}

\altaffiltext{2}{National Astronomical Observatories, CAS, Beijing 100012, 
P.R. China}
\altaffiltext{3}{Yunnan Observatory, CAS, Kunming 650011, P.R. China}

\begin{abstract}
{In this {\it Letter}, we make an attempt to estimate the masses of black 
holes and their distributions in BL Lac objects. We find there is a bimodal 
population of black hole masses in BL Lacs: the first has a lower-mass peak 
centred at $\sim 7.0 \times 10^6\sunm$ composed of high frequency-peaked BL 
Lacs (HBLs) and the second is a higher-mass ($>10^9\sunm$) population composed 
of low frequency-peaked BL Lacs (LBLs). The low mass family
of black holes in HBLs does not satisfy the Magorrian et al. relation
whereas the high mass family in LBLs does this relation and is in agreement
with Laor's limit. The origination of the bimodal distribution of black hole
mass remains open.}

\end{abstract}

\section{introduction}
The black hole masses and their distributions in active and inactive 
galaxies are of great interest in astrophysics (Richstone et al. 1998,
Kormendy \& Gebhardt 2001, Merritt \& Ferrarese 2001). BL 
Lacertae objects are characterized by their powerful featureless continuum 
radiation spanning from radio to gamma-ray bands (Bregman 1990). They play 
an important role in the evolution and unification scheme of active galactic 
nuclei (Urry \& Padovani 1995, Antonucci 1993).
With extensive studies of the BL Lacs, there has been considerable progress 
in understanding the properties of their multiwaveband continuum and radiation 
mechanisms. It is generally believed that synchrotron self-Compton (SSC) 
emission of high energy 
electrons in a relativistic jet is responsible for the multiwaveband continuum 
of the BL Lacs (Urry \& Padovani 1995). The emission is highly beamed. 
The featureless continuum of the BL Lacs only permits us to use timescale of 
variability $\Delta t$ to estimate black hole mass, $\bhm\le c^3\cd\Delta t/G$, 
where  $c$ is the speed of light, 
$G$ the gravitational constant and $\cd$ Doppler factor.
However this only provides a very crude upper limit of a black hole mass since
it is very difficult to determine the shortest timescale, 
especially when there is an uncertainty of relativistic boosting factor.
The host galaxies of BL Lacs have been discovered by HST, showing they are
normal giant ellipticals (Scarpa et al 2000, Urry et al. 2000). Urry et al. 
(2000) argue that the host galaxy can not determine the central engine
of BL Lac objects because there is no differences between HBLs and LBLs.
There is no measurements of black hole masses in BL Lac objects from stellar
dynamical method.

It has been pointed out that the masses of black holes in all the radio-loud 
quasars of low redshift PG sample  should exceed $10^9\sunm$ (Laor 2000). 
This suggests the black hole mass controls the radio-loudness $R$,  defined 
as the ratio of  radio to optical fluxes. The relativistic jet is generally
thought to originate from the innermost vicinity of black hole (Blandford 2001). 
The emission from the jet may reflect the black hole parameters in a more direct 
way. There is a large sample of BL Lacs that the peak frequencies and luminosities
are measured. This allows us to study the masses of black holes in BL Lacs.

In this {\it Letter}, we present a way to link the black hole mass to the
peak frequency and luminosity in BL Lac objects. This method allows us to
statistically study the mass distribution of black holes in a large sample.
We find there is a bimodal population of black hole
masses in BL Lac objects, corresponding to HBLs and LBLs, respectively. 
The implications of this bimodal distribution and 
Magorrian et al. relation are discussed. 

\section{Limits on supermassive black hole masses}
We intend to connect the mass of central black hole with the two available 
observables, the peak frequency $\nupk$ and its luminosity $\lpk$, rather
than to fit the continuum. As the first step, we express the Doppler factor 
as a function of $\nupk$ and $\lpk$. For the  simplicity, we assume a spherical 
geometry for the radiation region and a monoenergetic distribution of 
relativistic electrons responsible for the continuum emission, namely, 
$n(\ga)=n_0\delta(\ga-\ga_0)$, where $n_0$ is the electron density and $\ga_0$ 
the Lorentz factor of electrons. Using the standard formula for synchrotron 
emission in Pacholczyk (1970), $\epsilon_{\nu^{\prime}}=n_0c_3B\cf(x)$, where 
$c_3=1.87\times 10^{-23}$, we have the peak luminosity $\lpk$ in an observer's 
frame
\begin{equation}
\lpk=\cd^4 \frac{4}{3}\pi R_0^3\npkp \epsilon_{\npkp},
\end{equation}
where $\cf(x)=x\int_x^{\infty}K_{5/3}(z)dz$, $K_{5/3}(z)$ is the modified Bessel 
function with of order 5/3, $x=\nu^{\prime}/\nu^{\prime}_c$ and 
$\nu^{\prime}_c=4.2\times 10^6 B\ga_0^2$. The primed parameters
are in the co-moving frame of the jet. The peak frequency is given by 
$d\left[x\cf(x)\right]/dx=0$, namely, 
$\xpk=\left[2\cf(\xpk)/K_{5/3}(\xpk)\right]^{1/2}$, we have
\begin{equation}
\nupk=\cd \npkp=\cd \xpk \nu^{\prime}_c=5.5\times 10^6 \cd B\ga_0^2,
\end{equation}
where $\xpk\approx 1.32$. The radiating region must be optically thin for 
synchrotron self-absorption at the peak frequency, namely,
\begin{equation}
\tau_{\rm syn}(\nu^{\prime})=R_0c_4B^{3/2}n_0{\nu^{\prime}_c}^{-5/2}
       K_{5/3}(\xpk)=\tau_0,
\end{equation}
where $c_4=4.2\times 10^7$. The synchrotron self-absorption optical
depth can be obtained 
$\tau_0=K_{5/3}(\xpk)/K_{5/3}(x_{\rm sa})\approx (\nu_{\rm sa}/\nupk)^{5/3}$,
where $\nu_{\rm sa}$ is the self-absorption frequency. 
Following Ghisellini et al. (1993), we take the synchrotron self-absorption 
$\nu_{\rm sa}=3.0$GHz in BL Lacs. The radiation intensity, in the co-moving frame, 
is given by, $I_{\nu^{\prime}}^{\prime}=n_0c_3 B\cf(x) R_0$,
which is assumed to be homogeneous and isotropic. The energy density of
radiation field can be written as $U_{\rm rad}^{\prime}=4\pi c^{-1}\int_0^{\infty}
I_{\nu^{\prime}}^{\prime} d\nu^{\prime}\approx 4\pi c^{-1}c_3 n_0BR_0\nu_c^{\prime}$,
where we have used $F(x)\approx \delta (x-x_0)$ and $x_0=0.29$. 
The avoidance of Compton catastrophe leads to a condition
\begin{equation}
U_{\rm rad}^{\prime}=\chii U_{\rm B}^{\prime},
\end{equation}
where $\chi_{_1}\le 1$, and
the magnetic field energy density, $U_{\rm B}^{\prime}=B^2/8\pi$. In fact 
the condition $\chi_{_1}=1$ is generally satisfied in most BL Lacs. Lastly, 
we assume 
parameter $\chi_{_2}$ represents a departure from
equipartition of energy between the magnetic field and electrons,
\begin{equation} 
n_0\ga_0m_ec^2=\chiii U_{\rm B}^{\prime}.
\end{equation}
$\chi_{_2}=1$ means the equipartition between electrons and magnetic field.
We are not sure the value of $\chi_{_2}$, but we can constrain it from the
general consideration of unified scheme of BL Lacs.
Combining the above equations we have the Doppler factor as
\begin{equation}
\cd=12.2\nu_{\rm pk, 15}^{13/48}L_{\rm pk, 44}^{5/32}\Xi^{1/4},
\end{equation}
where $\nu_{\rm pk,15}=\nupk/10^{15}$Hz, 
$L_{\rm pk, 44}=\lpk/10^{44}$erg~s$^{-1}$,
$\Xi=\chii^{-17/8}\chiii^{5/4}\nu_{\rm sa,3}^{5/12}$, and
$\nu_{\rm sa,3}=\nu_{\rm sa}/$3GHz, and typically 
$\cd$=12 for $\Xi=1$.
The uncertain parameters $\chii$ and $\chiii$ are absorbed into $\Xi$ as
well as $\nu_{\rm sa}$,
which may be different in HBLs and LBLs. According to the unified
scheme of BL Lacs (Celotti et al. 1993, Urry \& Padovani 1995), typically,
the mean Doppler factor in HBLs is $\langle{\cd}_{\rm H}\rangle=6.0$ and 
$\langle\cd_{\rm L}\rangle=10$ in LBLs. From this argument, we obtain 
$\langle{\Xi}_{\rm H}\rangle=2.2\times 10^{-4}$ and 
$\langle{\Xi}_{\rm L}\rangle=0.74$ from the sample of Costamante et al. (2001). 
If $\chii=1$, namely the presence of equipartition between magnetic and radiation 
fields, we have $\langle\chiii^{\rm H}\rangle=1.2 \times 10^{-3}$ and 
$\langle\chiii^{\rm L}\rangle=0.8$.  This strongly implies that the energy 
density of electrons in HBLs is far away from equipartition with magnetic 
field while this equipartition generally holds in LBLs. 
This may reflect the reason why the peak frequency in HBLs
are more highly variable than that in LBLs (Costamante et al.
2001), because all likely mechanisms of electron acceleration
relate to magnetic field.  We find the dimension 
$R_0\propto \nupk^{-13/8}\lpk^{13/32}\chii^{7/32}\chiii^{-3/16}$,
and the number density of electrons 
$n_0\propto \nupk^{15/8}\lpk^{-15/32}\chii^{-19/32}\chiii^{1/16}$.

The ratio of kinetic to radiative (in comoving frame) luminosity, $\epsilon_{\rm r}$, 
can be expressed by 
$\epsilon_{\rm r}=L_{\rm kin}/L^{\prime}_{\rm rad}=\cd^4L_{\rm kin}/\lpk$, where 
$L^{\prime}_{\rm rad}$ is the radiative luminosity in jet comoving frame and 
$L_{\rm kin}=\pi R_0^2n_0m_pc^3\beta\Gamma^2$ (Celotti \& Fabian 1993), here $\beta$, 
$\Gamma$ and $m_p$ are the velocity, Lorentz factor and proton mass, respectively 
(we assume jet is proton-electron 
plasma). The number density of electrons, dimension of region and Doppler factor 
are already expressed by the peak frequency and luminosity, we have 
$\epsilon_{\rm r}\propto \nupk^{1/4}\lpk^{9/32}\chii^{-107/32}\chiii^{25/16}$.  
We then get
\begin{equation}
\frac{\pehp}{\pelp}= 
 \left(\frac{\langle\nupk^{\rm H}\rangle}{\langle\nupk^{\rm L}\rangle}\right)^{1/4}
 \left(\frac{\langle\lpk^{\rm H}\rangle}{\langle\lpk^{\rm L}\rangle}\right)^{9/32}
 \left(\frac{\langle{\chi^{\rm H}_2\rangle}}{\langle{\chi^{\rm L}_2\rangle}}\right)^{25/16},
\end{equation}
where we take $\chii=1$. We have $\pehp=1.5\times 10^{-4}\pelp$,
which is obtained from the sample of Costamante et al. (2001).
Interestingly the derived ratio of kinetic and radiative luminosity is roughly
consistent with the results in Celotti \& Fabian (1993), showing a bimodal distribution
of this ratio.

It was originally suggested that an optically thin torus powers the central 
engine of BL Lac (Rees et al. 1982), as stressed recently by Cavaliere
\& D'Elia (2001) in their model of blazar main sequence. 
The powerful emission from the relativistic jet originates from the efficient 
extraction of energy from the black hole spin by the Blandford-Zanjek 
process. Although it has been shown that this process in geometrically thin disk
is not so efficient to pump the energy from black hole spin (Livio, Ogilvie \&
Pringle 1999), it may dominate over the emission from disk in the ADAF regime.
With the self-similar solution of the optically thin ADAF of Narayan \& Yi (1994), 
the output power of the BZ process is  
\begin{equation}
\lbz=\epsilon_{\rm BZ}\dotm c^2,~~~{\rm with}~~~
                      \epsilon_{\rm BZ}\approx 1.8\times 10^{-2},
\end{equation} 
where $\dotm$ is the accretion 
rate of a black hole with maximum spin (Armitage \& Natarajan 1999). 
The ADAF dominates when the accretion rate is below the critical value
$\dotm_{\rm c}\le \alpha^2\dotm_{\rm Edd}$, where $\alpha$ is viscosity parameter and
the Eddington accretion rate is
$\dotm_{\rm Edd}=4\pi G\bhm/\eta \kes c=1.64\times 10^{26}M_{\rm BH,8}$(g/s), 
the electron scattering opacity $\kes=0.34$, $\eta=0.1$ and 
$M_{\rm BH,8}=M_{\rm BH}/10^8\sunm$ (Narayan \& Yi 1995). This condition works within a 
few hundred Schwartzchild radii for massive black holes, for which it is
independent of the black hole mass (Narayan \& Yi 1995).  According to the model
of their main sequence, the blazars' phenomenon can be understood in terms of different
levels of $\dot{m}$. The accretion rate $\dot{m}$ in BL Lac should be of $10^{-(3\sim 4)}$
(Cavaliere \& D'Elia 2001) (here we use different definition of Eddington limit). 
For all the BL Lacs,
it may be a good approximation that the critical value of the dimensionless 
accretion rate may be the
same $\dot{m}\approx \alpha^2$. 
This condition then yields a lower limit of black hole mass via
$L^{\prime}_{\rm rad}=L_{\rm BZ}/\epsilon_{\rm r}$. 
\begin{equation}
\bhm^{\rm ADAF} 
     \ge 
     \left\{\begin{array}{ll}
     5.68\times 10^6~\xi\nu^{-13/12}_{\rm pk,16}L^{3/8}_{\rm pk, 45}\sunm&{\rm (HBLs),}\\
     3.22\times 10^9~\xi\nu^{-13/12}_{\rm pk,13}L^{3/8}_{\rm pk, 46}\sunm&{\rm (LBLs),}
     \end{array}
     \right. 
\end{equation}
where we have used $\lpk=\cd^4L^{\prime}_{\rm rad}$ (e.g. Sikora et al. 1997), and 
$\xi=\langle\epsilon^{\rm L}_{\rm r}\rangle_{5}/\alpha^2_{-2}$. 
Here $\alpha_{-2}=\alpha/0.01$, and 
$\langle\epsilon^{\rm L}_{\rm r}\rangle_{5}=
\langle\epsilon_{\rm r}^{\rm L}\rangle/2.5\times 10^5$. $\alpha=0.01$ is 
reasonable for supporting ion tori (Rees et al. 1982) since a high $\alpha$ 
ADAF does not possess empty funnels (Narayan et al. 1998).
It should be addressed that the equation (9) is mainly based on the blazar main sequence
suggested by Cavaliere \& D'Elia (2001), namely, 
$\dot{m}\approx \alpha^2=10^{-(3\sim 4)}$,
which should be satisfied for BL Lac objects as a special group of blazar according to
the accretion level. 

On the other hand, Bondi accretion rate sets an upper limit of supermassive black hole mass.
{\it Chandra}'s observation shows that three giant elliptical galaxies, NGC 1399,
NGC 4472 and GNC 4636, have far below Bondi
accretion rate using the observed hot interstellar gas surface brightness profiles
(Loewenstein et al. 2001).  This strongly hints that the activity of giant elliptical 
galaxies may be due to the level of accretion rate, namely the activity condition is
$\dot{m}\ge\dot{m}_{\rm Bondi}$.
The dimensionless accretion rate given by equation (9) is
$\dot{m}=6.88\times 10^{-4}\nu_{\rm pk,16}^{-13/12}L_{\rm pk,45}^{3/8}M_{\rm BH,7}^{-1}$,
where we take $\xi=1$.  The Bondi accretion rate
in giant elliptical galaxies is 
\begin{equation}
\dot{m}_{\rm Bondi}=1.1\times 10^{-4} M_{\rm BH,9}n_{-1}c_{\rm s,5}^{-3},
\end{equation}
where $n_{-1}=n/0.1{\rm cm}^{-3}$ is the number density of the interstellar medium and
$c_{\rm s,5}=c_{\rm s}/500{\rm Km~s^{-1}}$ is the sound speed (Di Matteo \& Fabian 
1997).  The upper limit of black hole mass is given by the condition
$\dot{m}\ge\dot{m}_{\rm Bondi}$, 
\begin{equation}
M_{\rm BH}^{\rm Bondi}
             \le 1.5\times 10^8\nu_{\rm pk,15}^{-13/24}L_{\rm pk, 45}^{3/16}
             n_{-1}^{-1/2}c_{\rm s,5}^{3/2}~\sunm.
\end{equation}
The black hole mass is then confined by
\begin{equation}
M_{\rm BH}^{\rm ADAF}\le M_{\rm BH}\le M_{\rm BH}^{\rm Bondi}.
\end{equation}
The physical meanings of equations (9), (11) and (12) reflect the intrinsic
relation among the central engine, the emission and even its host galaxy, 
which allows us to deduce the mass of black hole.



\section{Masses of black holes in BL Lacs}
There is a largest sample including the high and low frequency-peaked BL 
Lacs composed of 1 Jy and Slew Survey BL Lacs, in which the peak 
frequency and its luminosity are given by fitting the SED with a 
polynomial (Costamante et al. 2001). There are three exceptional objects, Mrk 501, 
1ES 2344+514 and 1ES 1426+428 that show highly variable
properties. Their peak frequencies change by as much as $10^{19}$Hz to 
$10^{16}$Hz and peak luminosities from $10^{45.2}$ to $10^{44.3}$ erg/s. We 
exclude the data of these three objects in their high states to leave
100 data. All the data can be obtained from Costamante (private communication). 

Figure 1 shows the distributions of upper and lower limit of black hole masses
in BL Lac objects. 
The uncertainty in eq.(9) is due to the parameter $\xi$, namely,
$\alpha$ and $\langle\epsilon_{\rm r}^{\rm L}\rangle$. 
However we can use their typical values, $\alpha=0.01$ and 
$\langle\epsilon_{\rm r}^{\rm L}\rangle=2.5\times 10^5$, then $\xi=1$.
We find the high mass peak is in good agreement with Laor's limit,
which he find the mass of black hole in radio-loud quasar should 
exceed $10^9\sunm$ (Laor 2000). Thus we think $\xi=1$ is
reasonable. 
\vskip 2mm

\figurenum{1}
\centerline{\includegraphics[angle=-90,width=8.5cm]{mass_dis.ps}}
\figcaption{\footnotesize 
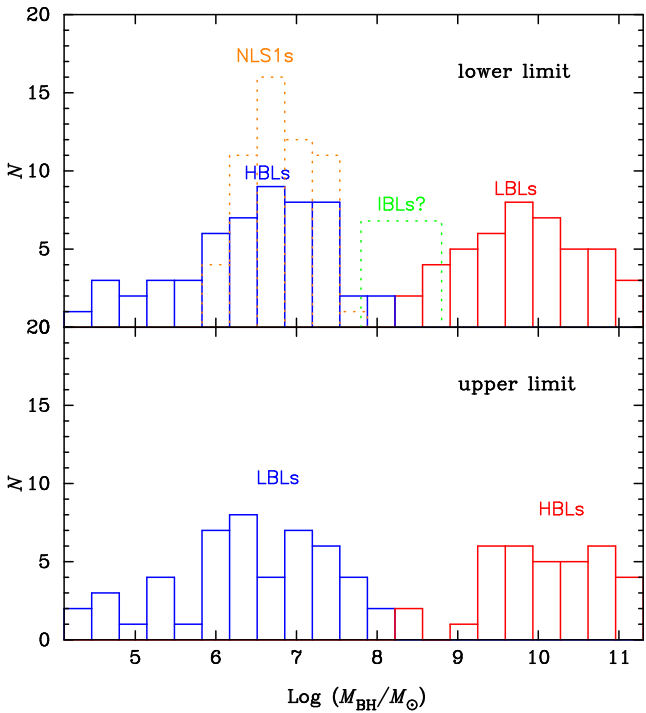
The mass distribution of black holes in BL Lac objects (solid line).
The high ($\nupk\ge 10^{15}$Hz) and low ($\nupk\le 10^{15}$Hz)
frequency-peaked BL Lacs are usually referred to as blue and red BL Lacs, 
respectively. The upper pannel is the lower limit mass distribution whereas
the low pannel is the upper limit mass distribution.
There is a bimodal mass population of black holes in HBLs and LBLs. 
It is very interesting that the mass distribution of black holes in narrow line 
Seyfert 1 galaxies (dotted line) overlaps that of high frequency-peaked BL Lac 
objects. 
\label{fig1}}
\centerline{}

It is interesting to find there is evidently a low peak of black hole mass 
populations centered roughly at $7.0\times 10^6\sunm$ from 
$(1.0\sim 30)\times 10^6\sunm$ when 
the high peak reconcile with the Laor's limit. The mass of black hole in
Mrk 421 has been estimated to be $\approx 10^7\sunm$ with accretion rate
$\dot{m}\sim 10^{-3}$(Celotti, Fabian \& Rees 1998). The present
result overlaps this mass. 

Many intermediate BL Lacs have been found in the RGB (Laurent-Muehleiser et al.
1998), DXRBS (Perlman et al. 1998) and REX surveys (Caccianiga et al. 1999). 
Simulations show that the IBL content, even the deeper ones, 
is affected more strongly by the sample flux than by intrinsic 
properties \cite{gi01}. We have 
not an homogeneous complete sample composed of HBLs, IBLs and LBLs. It is still 
unknown the unbiased occurring of BL Lacs according to peak frequency $\nupk$. 
It may have two possible cases. First if the IBL fraction to the total of BL Lacs
is still smaller than that of HBL and LBL, the bimodal
distribution of BL Lacs still holds. Second, this fraction is comparable with
the other two, then the gap between the two peaks will be filled up. 
There is no available data of peak frequency
and peak luminosity for a sample of intermediate BL Lacs, 
but we presume an arbitrary distribution of the second case of IBL for
illustration in Figure 1.

\section{black holes and spheroids in BL Lac objects}
The black hole mass is generally thought to increase with the mass of spheroid
in galaxies (Magorrian et al 1998) and quasars (Laor 1998), especially the mass 
is strongly correlated with dispersion velocity (Ferrarese \& Merritt 2000, 
Gebhardt et al 2000). The high mass family of black holes in LBLs roughly agrees 
with the Magorrian et al.  relation.  

With the mean absolute magnitude of X-ray selected BL Lacs at $R$ band
$\langle M_{\rm R}\rangle=-23.9$ (Falomo et al. 2000), we get the mean mass of 
black holes in HBL $\langle M_{\rm BH}^{\rm HBL}\rangle=10^9\sunm$
if the relation $\log(M_{\rm BH}/M_{\odot})=-0.51M_{\rm R}-3.2$
(McLure \& Dunlop 2001) works. 
Clearly, the low mass family consisting of HBLs does not satisfy 
the Magorrian et al. relation. 
If the mass of black hole is of $10^9\sunm$ in HBLs,
the dimensionless accretion rate from equation (9) is
$\dot{m}=6.88\times 10^{-6}\nu_{\rm pk,16}^{-13/12}L_{\rm pk,45}^{3/8}M_{\rm BH,9}^{-1}$,
We should note that:
1) Such a low accretion rate breaks down the blazar's main sequence according to
accretion level (Cavaliere \& D'Elia 2001). 2) We find 
$\dot{m}\ll \dot{m}_{\rm Bondi}$ (from eq.10) if the mean mass of black holes in 
HBLs is $\langle M_{\rm BH}\rangle=10^9\sunm$. If the accretion rate in HBLs is 
indeed far below Bondi accretion rate, this means that {\it all} the face-on giant 
elliptical galaxies will appears as HBLs. If the maximum viewing angle is 
$\theta_{\rm max}$, the appearance frequence of BL Lac objects will be given by 
$f=(1-\cos\theta_{\rm max})/2\approx 6.7\%$ with $\theta_{\rm max}=30^{\circ}$ 
(Urry \& Padovani 1995) based on random orientation of giant elliptical galaxies. 
The number density of BL Lacs will be given by $N_{\rm BL}=fN_{\rm GE}$, where 
$N_{\rm GE}$ is the elliptical galaxy number density. 
We get the number density of giant elliptical galaxies is conservatively 
$N_{\rm GE}\sim 10^4~{\rm Gpc^{-3}}$ for $M_V<-23.0$ in the local space (e.g. 
Table 3 in Woltjer 1990).
The predicted number density of BL Lacs based
on the giant elliptical galaxies will be $fN_{\rm GE}\sim 600~{\rm Gpc^{-3}}$. 
However, the observed number density of BL Lacs is only
$N_{\rm BL}\le 1.0~{\rm Gpc^{-3}}$ estimated from the luminosity function 
in their Fig.18 of Urry \& Padovani (1995). Evidently the predicted number density 
is much higher than the observed by a factor of $\ge 600$!
 To iron out this inconsistency 
is to lower the mass of black holes in HBLs. These arguments suggest that 
the black hole mass might not obey the Magorrian et al. relation
and support the blazar main sequence according to the accretion level.

There several narrow line Seyfert 1 galaxies as radio-loud objects, which are
PKS 2004-447 ($R>1700$) of 5$\times 10^6\sunm$ (Oshlack, Webster \& Whiting 2001),
PKS 0558-504 ($R=27$), RGB J0044+193
($R=31$) (Siebert et al. 1999) and RX J0134.2-4258
($R=71$) (Grupe et al. 2000). These objects evidently comprise a special
population in the plot of radio loudness and black hole mass (Laor 2000).
The distinguished features of narrow line Seyfert 1 galaxies (NLS1s) can be
explained by the hypothesis that they contain less massive black holes with
relatively high accretion rates, close to or even super-Eddington, and disks
which are face-on to the observer (Boller, Brandt \& Fink 1996) as BL Lacs. 
A heterogeneous sample of 59 NLS1 galaxies with good spectroscope observations 
is provided (Veron-Cetty, Veron \& Goncalves 2001). The masses of black holes 
in NLS1s has be obtained from their \hb width (Wang \& Lu 2001), we plot 
the mass distribution in Figure 1. It is found that the mass 
distribution of black holes in high frequency-peaked BL Lacs nicely overlaps that 
of the narrow line Seyfert 1 galaxies. We thus suggest that there might be an
evolutionary relation from a BL Lac to a narrow line Seyfert 1 nucleus
although their host galaxies are quite different. 

\section{Conclusions and Discussions}
This study shows that there are two populations of black holes in
BL Lacs corresponding to the low mass family of HBLs and the high mass
family of LBLs. The high mass family 
satisfies the Magorrian et al. relation roughly and is in agreement with Laor's
limit whereas the low mass family may not obey the Magorrian et al relation. 

Based on the unified scheme of HBLs and LBLs, we find that the energy density of
electrons is far below the equipartition with magnetic field in HBLs whereas this 
equipartition roughly holds in LBLs.
This reflects why the peak frequencies in HBLs are usually highly variable.

The bimodal distribution of black holes sets strong constraints on the model of
evolution
of BL Lacs (Ghisellini et al. 1998, B\"ottcher \& Dermer 2001). First if accretion 
rate of black holes, $\dot{m}\approx \alpha^2$, holds in HBLs constantly, it will 
take more than $10^{11}$yrs
for black holes to grow up to be the same as in LBLs. In such a case it seems that
there is no evolutionary connection between HBLs and LBLs. On the other hand, if
the model of multiple generation of short-lived quasars works (Haehnelt et al 1998,
Kuhn et al 2001), the accretion rate increases and decreases around its peak
(Cavaliere \& Vittorini 1998), and then repeats in the life.
HBLs are only short-lived objects, there might be evolutionary sequence from 
HBLs$\rightarrow$LBLs. The possible intermediate objects are narrow line Seyfert 1
galaxies, which are undergoing very fast accretion.
The mass distribution of black holes in high frequency-peaked BL Lac 
objects is similar to that of the narrow line Seyfert 1 galaxies, likely
implying some evolutionary connections between the two kinds of galactic nuclei. 

The most uncertainty is the parameter $\xi$ due to $\alpha$ and $\epsilon_{\rm r}$, 
but the present results can not be modified since $\xi$ can not changed three orders.
On the other hand $\xi$ can be constrained by the Laor's limit, which makes the 
low mass family safe. It is generally thought the relativistic jet is originated
in the vicinity of central black hole. We would like to stress here that the emission
from jet may directly take the most important information of black hole parameters.
The present method is robust in the view of this point. Although we use the 
monoenergetic distribution of relativistic electrons to calculate the emission from 
jet, the present method works as a robust tool to extract statistic properties of 
BL Lacs. It highly desired to measure the mass of black holes via stellar dynamics 
or gas dynamical methods in centers of BL Lacs to confirm the present results.
The origination of the bimodal distribution of black hole masses in BL Lacs 
remains open.

\acknowledgements
{J.M.W. is very grateful to an anonymous referee for his constructive comments.
C.A. Tout is acknowledged for his careful reading of the
manuscript and comments. Helpful discussions with A. Laor, H. Netzer,
M. B\"ottcher, A. Cavaliere,
X. Cao and Y.Y. Zhou are appreciated. J.M.W. is supported by ``Hundred Talents 
Program of CAS". This research is financed by the Special Funds for Major 
State Basic Research Project and NSFC.}

\end{document}